\documentclass[preprintnumbers,amsmath,amssymbm,prd]{revtex4}

\usepackage{epsfig}
\usepackage{graphicx}
\usepackage{amssymb}

\begin{document}
\title{Lower bound on the radii of black-hole shadows}
\author{Shahar Hod}
\affiliation{The Ruppin Academic Center, Emeq Hefer 40250, Israel}
\affiliation{ }
\affiliation{The Jerusalem Multidisciplinary Institute, Jerusalem 91010, Israel}
\date{\today}

\begin{abstract}
\ \ \ The non-linearly coupled Einstein-matter field equations predict the existence of shadows with 
well-defined boundaries around black holes. 
We prove that, in spherically symmetric hairy black-hole spacetimes whose matter fields satisfy the weak energy condition, 
the radii of these shadows are bounded from below by the dimensionless relation  $r_{\text{sh}}/r_{\text{H}}\geq 3\sqrt{3}/2$, 
where $r_{\text{H}}$ is the horizon radius of the central hairy black hole. 
The characteristic shadow of the (bald) Schwarzschild black-hole spacetime saturates the analytically 
derived lower bound.
\end{abstract}
\bigskip
\maketitle

\section{Introduction}

The Event Horizon Telescope has recently provided the first images of the supermassive black holes 
M87* and Sgr A* \cite{EHT1,EHT2,EHT3,EHT4}. 
These groundbreaking observations enhance our ability to study the nontrivial physics of strong-gravity phenomena 
in the vicinity of black-hole event horizons.

Interestingly, the images released by the Event Horizon Telescope reveal fine structures that characterize the 
near-horizon strong-gravity region of the central black hole. 
In particular, the photon ring -- a closed null circular geodesic -- and the dark shadow around the 
central black hole, two fundamental features of highly curved spacetimes whose existence is predicted by 
the Einstein-matter field equations of general relativity, have been clearly observed by the 
Event Horizon Telescope \cite{EHT1,EHT2,EHT3,EHT4}. 

The shadow of a black hole, which is a two-dimensional dark region on the celestial sphere, 
represents the locus of light rays that intersect the photonsphere 
\cite{Bar,Chan,Shap,Herne,Hodns,YY,Mash,Goeb,Hod1,Dec,Hodhair,YP,Hodm,Hodub,Lu1,Ame,Ste} 
of the curved black-hole spacetime (and therefore the absorbing horizon of the central black hole). 
In particular, the radius $r_{\text{sh}}$ of the shadow as measured by asymptotic observers is closely related to 
the gravitationally lensed image and the radius $r_{\gamma}$ of the black-hole null circular geodesic [see Eq. (\ref{Eq19}) below].

Since the characteristic shadows of black-hole spacetimes represent a genuine strong-gravity feature 
of general relativity, they have attracted much attention from physicists and mathematicians 
over the past five decades. 
In particular, the shadow of the canonical Schwarzschild (vacuum) black-hole spacetime 
was first analyzed in \cite{Syn}, where it was shown that 
its radius satisfies the remarkably simple dimensionless relation
\begin{equation}\label{Eq1}
{{r^{\text{Sch}}_{\text{sh}}}\over{r_{\text{H}}}}={{3\sqrt{3}}\over{2}}\  ,
\end{equation}
where $r_{\text{H}}$ is the horizon radius of the central Schwarzschild black hole.  

The main goal of the present paper is to analyze the physical and mathematical properties of 
shadows in non-trivial (non-vacuum) black-hole spacetimes. 
In particular, motivated by the physical importance of shadows in curved black-hole spacetimes, in the 
present paper we shall address the following physically intriguing question: 
How close can the outer edge of a black-hole shadow be to the horizon of the corresponding central black hole? 

It is worth noting that, since the horizon of a black hole cannot be observed directly, it is of physical interest to derive 
in a mathematically consistent manner a lower bound on the ratio $r_{\text{sh}}/r_{\text{H}}$ that characterizes 
a black-hole spacetime. 
In particular, a lower bound on the dimensionless ratio $r_{\text{sh}}/r_{\text{H}}$ 
would provide an important observational tool for bounding from above the horizon radii of 
observed black holes.

Interestingly, below we shall prove that, under physically plausible conditions, 
the non-linearly coupled Einstein-matter field equations yield an explicit lower bound [see Eq. (\ref{Eq30}) below] 
on the dimensionless radii $r_{\text{sh}}/r_{\text{H}}$ of shadows in 
spherically symmetric non-vacuum (hairy) black-hole spacetimes. 

\section{Review of previous analytical results}

In a very important paper \cite{Ver}, Yang and L\"u recently used the Einstein equations for 
matter fields that satisfy both the null energy condition (NEC), the weak energy condition (WEC), 
and the strong energy condition (SEC) to prove the physically interesting lower bound
\begin{equation}\label{Eq2}
\big\{\text{NEC}\ \ \ \land\ \ \ \ \text{WEC}\ \ \ \land\ \ \ \ \text{SEC}\big\}
\ \ \Longrightarrow\ \ r_{\text{sh}}\geq \sqrt{3}\cdot r_{\gamma}\
\end{equation}
on the radii of black-hole shadows in spherically symmetric spacetimes, where 
$r_{\gamma}$ is the radius of the outermost null circular geodesic (closed light ring) that characterizes the 
corresponding black-hole spacetime. 

It is of physical interest to obtain a bound on the radius $r_{\text{sh}}$ of the black-hole shadow 
in terms of the horizon radius $r_{\text{H}}$ of the corresponding central black hole. 
To this end, we point out that it was shown in \cite{Ver} that, in certain cases, 
the radius of the black-hole light ring can be bounded from below by the relation 
\begin{equation}\label{Eq3}
r_{\gamma}\geq {3\over2}r_{\text{H}}\  .
\end{equation}
In particular, it was shown in \cite{Ver} that the inequality (\ref{Eq3}) holds true 
if there exists a radially dependent function $\Xi(r)$ such that 
\begin{equation}\label{Eq4}
\Big\{-\rho(r)\leq\Xi(r)\leq p(r)\ \ \ \ \text{and}\ \ \ \ {{d[r^2\Xi(r)]}\over{dr}}\geq0\Big\}
\ \ \Longrightarrow\ \ r_{\gamma}\geq {3\over2}r_{\text{H}}\  ,
\end{equation}
where $\rho(r)$ and $p(r)$ are respectively the energy density and the radial pressure of the hairy matter 
fields. 

In addition, it was shown in \cite{Hodlbd} that the special relation (\ref{Eq3}) 
characterizes hairy black-hole spacetimes for which the radial pressure function $P(r)\equiv |r^3p(r)|$ of 
the external matter fields decreases monotonically:
\begin{equation}\label{Eq5}
{{d|r^3p(r)|}\over{dr}}\leq0\ \ \Longrightarrow\ \ r_{\gamma}\geq {3\over2}r_{\text{H}}\  .
\end{equation}

Taking cognizance of Eqs. (\ref{Eq2}), (\ref{Eq4}), and (\ref{Eq5}), one concludes that, in 
certain {\it special} cases, the 
radius $r_{\text{sh}}$ of the black-hole shadow can be bounded from below 
explicitly in terms of the black-hole horizon radius $r_{\text{H}}$. 
In particular, one finds that spherically symmetric hairy black-hole spacetimes whose external matter fields 
satisfy the following set of four conditions [see Eqs. (\ref{Eq2}), (\ref{Eq4}), and (\ref{Eq5})]: 
\newline
(i) the null energy condition, 
\newline
(ii) the weak energy condition, 
\newline
(iii) the strong energy condition, 
\newline 
and 
\newline
(iv) condition (\ref{Eq4}) or condition (\ref{Eq5}), 
\newline
are characterized by the lower bound
\begin{equation}\label{Eq6}
\big\{\text{NEC}\ \ \ \land\ \ \ \ \text{WEC}\ \ \ \land\ \ \ \ \text{SEC}\ \ \ \land
\ \ \ \ \{(\ref{Eq4})\ \lor\ (\ref{Eq5})\}\big\}
\ \ \Longrightarrow\ \ r_{\text{sh}}\geq {{3\sqrt{3}}\over{2}}\cdot r_{\text{H}}\  .
\end{equation}

In the present compact paper we shall reveal the physically intriguing fact that the lower bound 
$r_{\text{sh}}\geq ({{3\sqrt{3}}/{2}})\cdot r_{\text{H}}$ 
on the radii of black-hole shadows, which is expressed solely in terms of the black-hole horizon 
radius $r_{\text{H}}$, is a {\it generic} feature of hairy black-hole spacetimes. 
In particular, below we shall explicitly prove that 
the non-linearly coupled Einstein-matter field equations, supplemented only by the {\it weak} energy condition (WEC), 
impose the same lower bound on the characteristic dimensionless radii $r_{\text{sh}}/r_{\text{H}}$ 
of black-hole shadows.  

\section{Description of the system}

We shall study, using analytical techniques, the radii of shadows that characterize non-trivial (non-vacuum) 
hairy black-hole spacetimes. 
The spherically symmetric spacetimes are described, using the Schwarzschild areal coordinates, by the curved line element
\cite{Hodlbd,Notesch2,Noteunit}
\begin{equation}\label{Eq7}
ds^2=-e^{-2\delta}\mu dt^2 +\mu^{-1}dr^2+r^2(d\theta^2 +\sin^2\theta d\phi^2)\  ,
\end{equation}
where $\mu=\mu(r)$ and $\delta=\delta(r)$ are radially-dependent 
dimensionless metric functions. 

Asymptotically flat spacetimes are characterized by the large-$r$ boundary conditions \cite{Bekreg}
\begin{equation}\label{Eq8}
\mu(r\to\infty)\to1\ 
\end{equation}
and 
\begin{equation}\label{Eq9}
\delta(r\to\infty)\to0\  .
\end{equation}
In addition, spatially regular horizons are characterized by the boundary conditions \cite{Bekreg}
\begin{equation}\label{Eq10}
\mu(r=r_{\text{H}})=0\
\end{equation}
and
\begin{equation}\label{Eq11}
\delta(r=r_{\text{H}})<\infty\ \ \ \ ; \ \ \ \ \Big[{{d\delta}\over{dr}}\Big]_{r=r_{\text{H}}}<\infty\
\end{equation}
at the radial location of the black-hole outermost horizon. 

The Einstein equations $G^{\mu}_{\nu}=8\pi T^{\mu}_{\nu}$ for the non-vacuum curved 
spacetime (\ref{Eq7}) yield the coupled differential equations \cite{Hodm}
\begin{equation}\label{Eq12}
{{d\mu}\over{dr}}=-8\pi r\rho+{{1-\mu}\over{r}}\
\end{equation}
and
\begin{equation}\label{Eq13}
{{d\delta}\over{dr}}=-{{4\pi r(\rho +p)}\over{\mu}}\  ,
\end{equation}
where \cite{Bond1}
\begin{equation}\label{Eq14}
\rho\equiv-T^{t}_{t}\ \ \ \ ,\ \ \ \ p\equiv T^{r}_{r}\ \ \ \ , \ \ \ \ p_T\equiv T^{\theta}_{\theta}=T^{\phi}_{\phi}\
\end{equation}
are respectively the energy density, the radial pressure, and the tangential pressure 
that characterize the external matter fields of the hairy black-hole spacetime (\ref{Eq7}). 
We consider hairy black-hole configurations whose external matter fields satisfy 
the weak energy condition \cite{Bekreg}:
\begin{equation}\label{Eq15}
\rho\geq0\ \ \ \ , \ \ \ \ \rho+p\geq0\  .
\end{equation}
 
The Einstein equation (\ref{Eq12}) yields the radially-dependent functional relation
\begin{equation}\label{Eq16}
\mu(r)=1-{{2m(r)}\over{r}}\  ,
\end{equation}
where 
\begin{equation}\label{Eq17}
m(r)=m(r_{\text{H}})+\int_{r_{\text{H}}}^{r} 4\pi r^{2}\rho(r)dr\
\end{equation}
is the gravitational mass contained within a sphere of radius $r\geq r_{\text{H}}$ with 
the simple horizon boundary relation [see Eqs. (\ref{Eq10}) and (\ref{Eq16})] 
\begin{equation}\label{Eq18}
m(r=r_{\text{H}})={{r_{\text{H}}}\over{2}}\  .
\end{equation}

\section{Lower bound on the radii of shadows in spherically symmetric hairy black-hole spacetimes}

In the present section we shall derive, using analytical techniques, a generic lower bound on the 
dimensionless ratio $r_{\text{sh}}/r_{\text{H}}$ that characterizes the 
dark shadows of spherically symmetric non-vacuum (hairy) black-hole spacetimes. 

To this end, we shall use the fact that the radius $r_{\text{sh}}$ of a black-hole shadow 
as observed by asymptotic observers is directly related to the gravitationally lensed image of 
the black-hole light ring (null circular geodesic) \cite{Pas}. 
In particular, in the curved black-hole spacetime (\ref{Eq7}) the relation between the two radii is given by \cite{Pas}
\begin{equation}\label{Eq19}
r_{\text{sh}}={{r_{\gamma}}\over{\sqrt{-g_{tt}(r_{\gamma})}}}\  ,
\end{equation}
where $r_{\gamma}$ is the radius of the light ring that characterizes the black-hole spcaetime. 
Taking cognizance of Eqs. (\ref{Eq7}), (\ref{Eq16}), and (\ref{Eq19}), one obtains the functional expression  
\begin{equation}\label{Eq20}
r_{\text{sh}}={{r_{\gamma} \cdot e^{\delta(r_{\gamma})}}\over
{\sqrt{1-{{2m(r_{\gamma})}\over{r_{\gamma}}}}}}\
\end{equation}
for the radius of the shadow that characterizes the spherically symmetric hairy black-hole spacetime (\ref{Eq7}). 

From Eqs. (\ref{Eq13}) and (\ref{Eq15}) one learns that the radially-dependent metric 
function $\delta(r)$ that characterizes the black-hole spacetime (\ref{Eq7}) 
is a monotonically decreasing function \cite{Noteas1}:
\begin{equation}\label{Eq21}
{{d\delta(r)}\over{dr}}\leq0\ \ \ \ \text{for}\ \ \ \ r\in[r_{\text{H}},\infty]\  .
\end{equation}
In addition, taking cognizance of Eqs. (\ref{Eq9}) and (\ref{Eq21}) one finds the relation \cite{Noteas2,DS1,DS2}
\begin{equation}\label{Eq22}
\delta(r)\geq0\ \ \ \ \text{for}\ \ \ \ r\in[r_{\text{H}},\infty]\  .
\end{equation}
Substituting the inequality (\ref{Eq22}) into the functional expression (\ref{Eq20}) 
for the radius of the shadow that characterizes the hairy black-hole spacetime (\ref{Eq7}), 
one obtains the lower bound
\begin{equation}\label{Eq23}
r_{\text{sh}}\geq{{r_{\gamma} }\over
{\sqrt{1-{{2m(r_{\gamma})}\over{r_{\gamma}}}}}}\  ,
\end{equation}
which yields the dimensionless inequality 
\begin{equation}\label{Eq24}
{{r_{\text{sh}}}\over{m(r_{\gamma})}}\geq
{\cal F}[m(r_{\gamma}),r_{\gamma}]\equiv
\Bigg[\sqrt{1-{{2m(r_{\gamma})}\over{r_{\gamma}}}}\cdot {{m(r_{\gamma})}\over{r_{\gamma}}}\Bigg]^{-1}\  .
\end{equation}

From Eq. (\ref{Eq24}) one deduces that, 
for a given value of the radius $r_{\gamma}$ that characterizes the null circular geodesic (closed light ring) of 
the hairy curved spacetime (\ref{Eq7}), the function ${\cal F}[m(r_{\gamma});r_{\gamma}]$ is minimized for
\begin{equation}\label{Eq25}
m(r_{\gamma})={1\over3}r_{\gamma}\  .
\end{equation}
In particular, taking cognizance of Eqs. (\ref{Eq24}) and (\ref{Eq25}) one obtains 
the universal ($r_{\gamma}$-{\it independent}) relation
\begin{equation}\label{Eq26}
\text{min}\big\{{\cal F}[m(r_{\gamma});r_{\gamma}]\big\}={\cal F}[m(r_{\gamma})=
r_{\gamma}/3;r_{\gamma}]=3\sqrt{3}\  ,
\end{equation}
which yields the dimensionless inequality
\begin{equation}\label{Eq27}
{{r_{\text{sh}}}\over{m(r_{\gamma})}}\geq
\Big[{{r_{\text{sh}}}\over{m(r_{\gamma})}}\Big]_{\text{min}}=3\sqrt{3}\  .
\end{equation} 

In addition, taking cognizance of Eqs. (\ref{Eq15}), (\ref{Eq17}), and (\ref{Eq18}), one finds the relation
\begin{equation}\label{Eq28}
m(r_{\gamma})\geq m(r_{\text{H}})={{r_{\text{H}}}\over{2}}\  .
\end{equation}
Substituting (\ref{Eq28}) into (\ref{Eq27}), one obtains the lower bound
\begin{equation}\label{Eq29}
r_{\text{sh}}\geq{{3\sqrt{3}}\over{2}}\cdot r_{\text{H}}\ 
\end{equation}
on the characteristic radii of shadows in spherically symmetric hairy black-hole configurations. 

It is worth emphasizing again that the physically important bound (\ref{Eq29}) 
was previously derived in the highly interesting paper \cite{Ver} for matter fields that satisfy 
both the null energy condition, the weak energy condition, the strong energy condition, and at least one 
of the conditions (\ref{Eq4}) or (\ref{Eq5}) [see Eq. (\ref{Eq6})]. 
In the present paper we have explicitly proved that assuming the weak energy condition alone 
is sufficient to derive the lower bound (\ref{Eq29}).

\section{Summary}

The recently published images of light rings and dark shadows around the supermassive black holes 
M87* and Sgr A* \cite{EHT1,EHT2,EHT3,EHT4} improve 
our ability to probe gravitational phenomena in the highly curved spacetime 
regions near black-hole horizons. 

Motivated by the important theoretical and observational roles played by 
shadows in curved black-hole spacetimes, in the present compact paper we have raised the following 
physically intriguing question: 
How close can the outer edge of a black-hole shadow be to the outermost horizon of the central black hole? 

In order to address this important question we have used the non-linearly coupled Einstein-matter field equations 
supplemented by the weak energy condition. In particular, we have derived the 
remarkably compact dimensionless inequality [see Eq. (\ref{Eq29})] 
\begin{equation}\label{Eq30}
{{r_{\text{sh}}}\over{r_{\text{H}}}}\geq{{3\sqrt{3}}\over{2}}\  .
\end{equation}
The lower bound (\ref{Eq30}) determines 
the smallest possible dimensionless shadow-to-horizon radius ratio in spherically symmetric 
non-trivial (non-vacuum) black-hole spacetimes 
whose external matter fields (hairy configurations) satisfy the physically motivated weak energy 
condition \cite{Notephy}.

It is important to note that, within the sector of asymptotically flat black-hole spacetimes \cite{Noteas2}, 
the analytically derived lower bound (\ref{Eq30}) is of general validity as physically acceptable matter and radiation fields 
satisfy the weak energy condition (\ref{Eq15}) employed in its derivation.

It is worth emphasizing the fact that, since a black-hole horizon cannot be observed directly, the analytically 
derived relation (\ref{Eq30}) provides an important observational tool for placing an upper bound on the horizon radii 
of observed black holes. 
In addition, it is physically interesting to point out that the lower bound (\ref{Eq30}) 
is saturated by the characteristic shadow of the (bald) Schwarzschild black-hole spacetime.


\bigskip
\noindent {\bf ACKNOWLEDGMENTS}

This research is supported by the Carmel Science Foundation. I thank
Yael Oren, Arbel M. Ongo, Ayelet B. Lata, and Alona B. Tea for
stimulating discussions.



\begin{thebibliography}{99}
\bibitem{Notesch2} Here $\{t,r,\theta,\phi\}$ are the Schwarzschild coordinates of 
the spherically symmetric curved black-hole spacetime.
\bibitem{Noteunit} We use gravitational units in which $G=c=1$.
\bibitem{Noteas1} It is important to emphasize that our theorem is restricted to hairy black-hole spacetimes 
whose energy-momentum tensors respect the weak energy condition $\rho+p\geq0$ [see Eq. (\ref{Eq15})] which implies 
the functional behavior (\ref{Eq21}). In particular, a violation of the weak energy condition may allow the 
existence of spacetime regions that are characterized by the relation $\delta(r)<0$, in which case the 
inequality (\ref{Eq23}), which is used to derive the lower bound on the dimensionless 
ratio ${{r_{\text{sh}}}/{r_{\text{H}}}}$ [see Eq. (\ref{Eq30}) below] may be violated.
\bibitem{Noteas2} It is important to emphasize that our theorem is formally valid for 
asymptotically flat black-hole spacetimes that are characterized by the relation (\ref{Eq9}). 
Nevertheless, it is of physical interest to note that, provided the weak energy condition (\ref{Eq15}) is satisfied, 
our proof is also valid for non-asymptotically flat black-hole spacetimes with a non-zero cosmological constant that 
respect the property $\delta(r_{\text{max}})\to0$ (see \cite{DS1,DS2} and references therein).
\bibitem{Notephy} It is worth noting that the existence of a lower bound on the radii of black-hole shadows 
has a simple physical reason. In particular, the shadow of a black hole is determined by the set of null geodesics 
that fail to reach a distant observer because they are captured by the spacetime curvature near the black-hole horizon. 
The non-linearly coupled Einstein-matter field equations relate the existence of an event horizon 
to a minimum curvature scale in the near-horizon region, which in turn implies a minimum deflection of light rays. 
As a consequence, for a given radius of the central black hole, there exists a smallest impact parameter 
for photons that can escape to infinity. This suggests the existence of a universal lower bound on the dimensionless 
radius ${{r_{\text{sh}}}/{r_{\text{H}}}}$ of a black-hole shadow. In the present compact paper we have 
used the Einstein-matter field equations to derive an explicit expression [see Eq. (\ref{Eq30})] 
for the lower bound on the dimensionless size ${{r_{\text{sh}}}/{r_{\text{H}}}}$ of a black-hole shadow.
\end{thebibliography}
\end{document}